# Accelerators Beyond The Tevatron?

Joseph Lach

*Fermi National Accelerator Laboratory*
*P.O Box 500, Batavia Illinois*

March 16, 2010

**Abstract.** Following the successful operation of the Fermilab superconducting accelerator three new higher energy accelerators were planned. They were the UNK in the Soviet Union, the LHC in Europe, and the SSC in the United States. All were expected to start producing physics about 1995. They did not. Why?

## JOURNEY: 6.2 BEV → 33 GEV → 200 → 400 → TEV

In the spring of 1963 I completed my thesis work at the Lawrence Radiation Laboratory (LRL) in Berkeley. My thesis used the 6.2 BeV accelerator – The Bevatron - then the world's highest energy proton accelerator. I then joined the Yale University Physics Department and was a builder of one of the first rf-separated beams[1] at the Alternation Gradient Synchrotron (AGS) at Brookhaven National Laboratory - then the highest energy (33 GeV) proton accelerator.

I joined the 1965 "summer study" at LRL to work on experimental beams for the LRL design of the proposed 200 BeV Accelerator. At the end of that summer, the LRL director, received a letter from an unknown – at least to me – physicist from a small university in upstate New York. In his scathing letter[2] Robert Wilson said their accelerator design was "much too conservative" and "lacking in imagination".

In the summer of 1969 I joined the National Accelerator Laboratory (NAL) in Batavia IL. Robert Wilson, now Director, had decided that the 200 BeV Accelerator would now be a 400 BeV accelerator. The construction schedule would be advanced by one year[2] and this would be done with a reduced construction cost!

The NAL accelerator achieved 200 GeV beam in March 1972 and a full scale research program at 400 GeV by 1975.

The NAL overview in Figure 1 shows the four mile circumference ring and experimental areas. Figure 2 shows the interior of the accelerator tunnel and the conventional magnets used for the program just described. But now one also sees the second lower ring of superconducting magnets that evolved into the Tevatron.

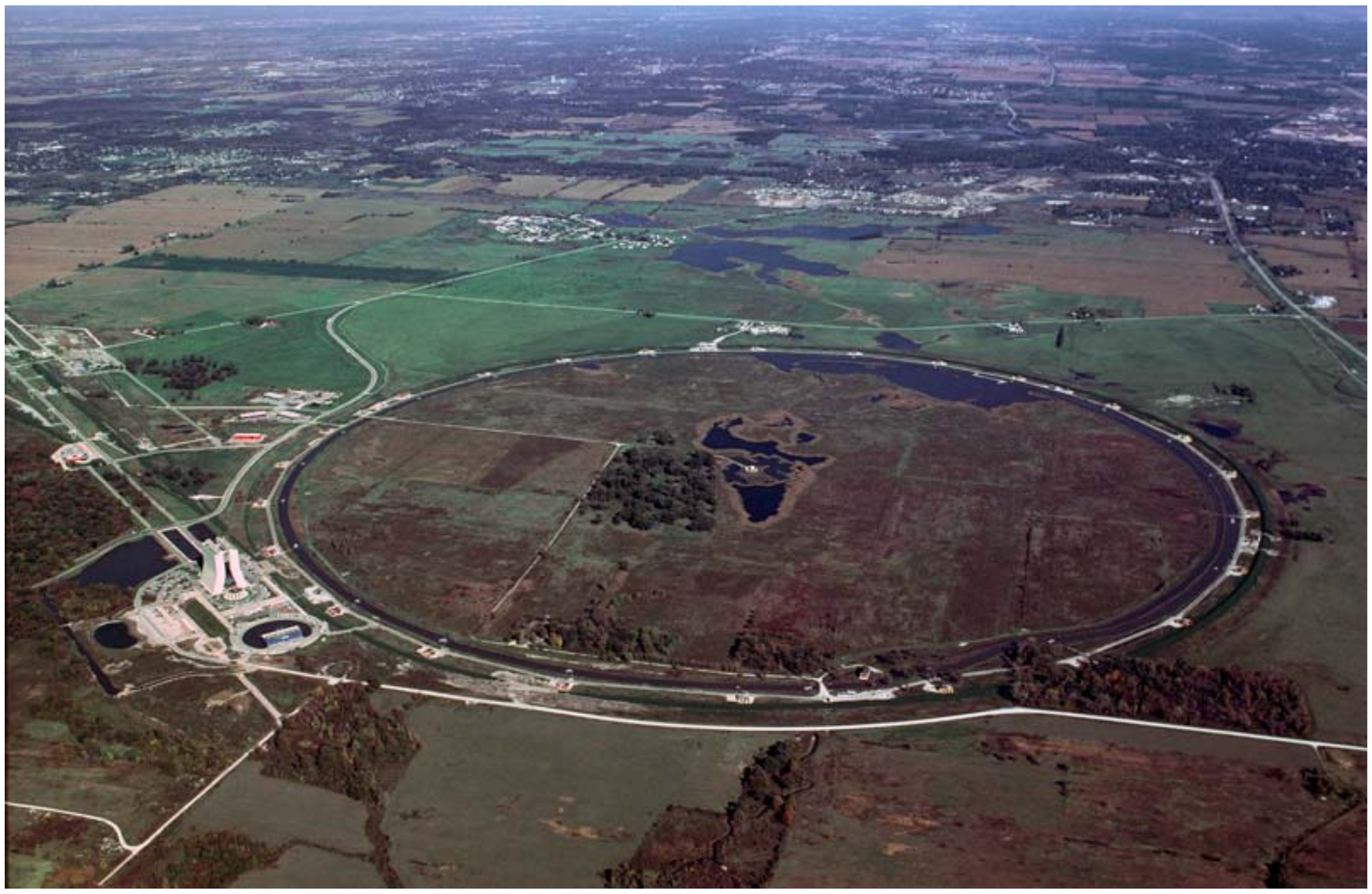

**Figure 1. The four mile ring contains the conventional 400 GeV accelerator.**

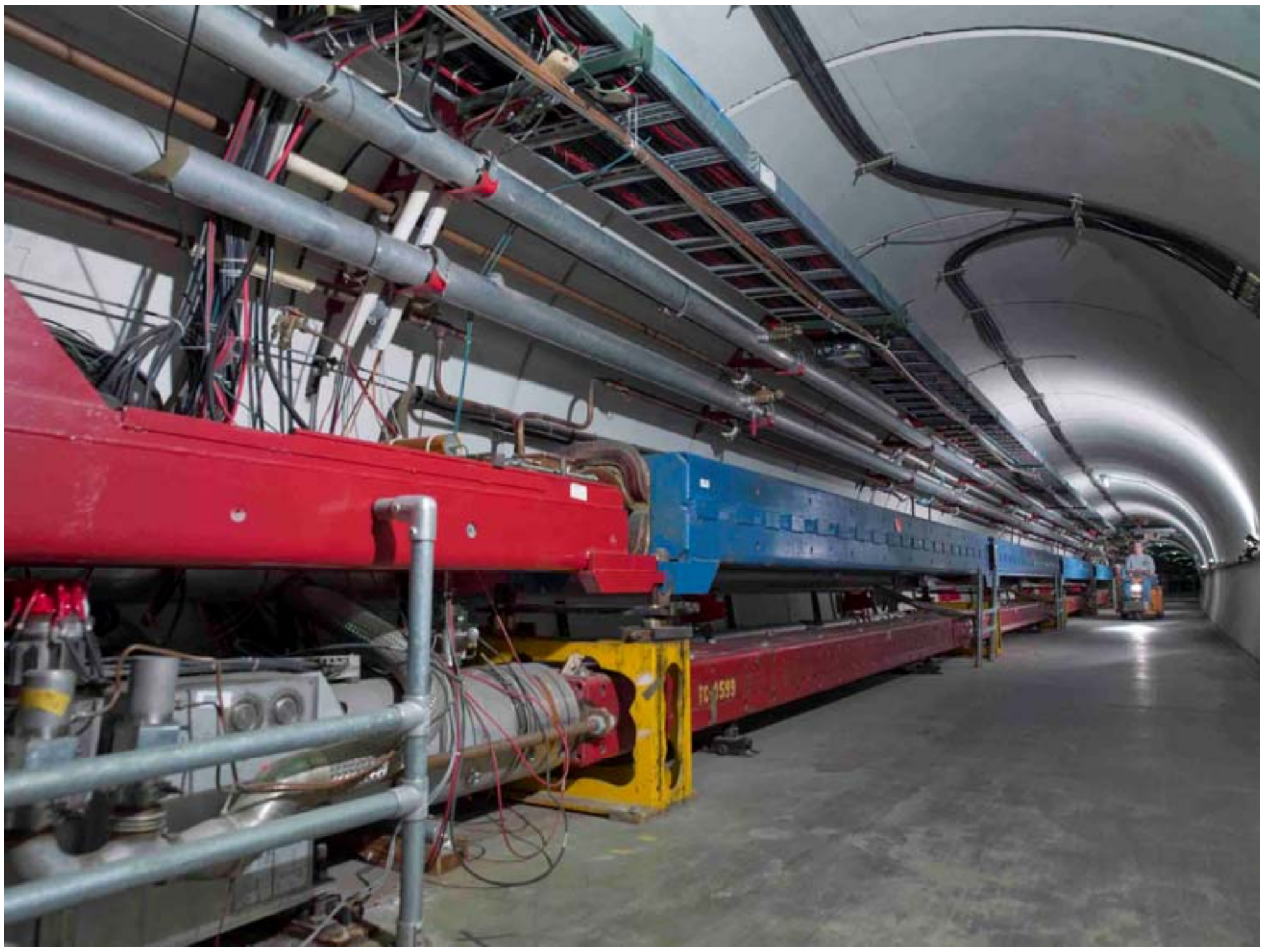

**Figure 2. Superconducting magnets below the conventional magnets**.

# THREE NEW ACCELERATORS

In 1986 Fermilab's high-energy superconducting proton accelerator (Tevatron) began operation at 400 GeV. In 1987 operation began at 800 GeV. This successful operation of a large superconducting accelerator energized the high-energy physics community into a new era of accelerator construction. The recorder of progress in this field, the Particle Data Group at LBL published the following table in their 1988 edition.

**HIGH-ENERGY COLLIDER PARAMETERS (Cont'd)**

**$pp$, $\bar{p}p$, and $ep$ Colliders**

The numbers here were received from representatives of each collider by late 1987. Numbers are subject to change, and many are only estimates. (Quantities are, where appropriate, r.m.s. $H$ ≡ horizontal direction, $V$ ≡ vertical direction, s.c. ≡ superconducting.)

| | $Sp\bar{p}S$ (CERN) | TEVATRON (Fermilab) | HERA (DESY) | UNK (Serpukhov) | LHC (CERN) | | SSC (USA) |
|---|---|---|---|---|---|---|---|
| Physics start date | 1981 | 1987 | Spring 1990 | 1995 ? | 1995 ? | | 1996 ? |
| Particles collided | $p\bar{p}$ | $p\bar{p}$ | $ep$ | $pp$ | $pp$ | $ep$ | $pp$ |
| Max. beam energy (TeV) | 0.315 (0.45 in pulsed mode) | 0.9−1.0 | $e$: 0.026<br>$p$: 0.82 | 3 | 8 | $e$: 0.05<br>$p$: 8 | 20 |
| Injection energy (TeV) | 0.026 | 0.15 | $e$: 0.014<br>$p$: 0.040 | 0.4 | 0.450 | $e$: 0.02<br>$p$: 0.450 | 1 |
| Luminosity ($10^{30}$ cm$^{-2}$ s$^{-1}$) | 0.3<br>3 (1988) | 0.5−1.0 | 15 | 400 | 1400 | 200 | 1000, $\beta^* = 0.5$ m<br>56, $\beta^* = 10$ m |
| Circumference (km) | 6.911 | 6.28 | 6.336 | 20.772 | 26.659 | | 83.631 |
| No. of interaction regions | 2 | 2 high $\mathcal{L}$<br>2 low $\mathcal{L}$ | 3 | 4 | 7 | 3 | 4<br>(initially) |
| No. of particles per bunch (units $10^{10}$) | $p$: 15<br>$\bar{p}$: 2−10 | $p$: 6<br>$\bar{p}$: 2 | $e$: 3.48<br>$p$: 10 | 3 | 2.5 | $e$: 8<br>$p$: 30 | 0.80 |
| No. of bunches per ring per species | 3−6 | 3−6 | 220 | 8,000 | 3564 | 540 | 15,456 |

**Figure 3. High-Energy Accelerator parameters 1988**

Note that three new accelerators appear in the above figure. They are the UNK being constructed at Serpukov in the Soviet Union, the Large Hadron Collider (LHC) in Europe at CERN, and the Superconducting Super Collider (SSC) at a not yet chosen U.S. site. The physics start date for each of these accelerators was estimated to be 1995-6. These were each major scientific projects with costs in billions of dollars.

The following table contains a time line for the new accelerator construction and some other relevant events.

| Accelerator Time Lines | | |
|---|---|---|
| Jul-83 | TM-1218 Tunneling Beyond Fermilab Site | Joe Lach |
| Jan-84 | Construction of UNK starts, 20.8 km ring | Victor Yarba |
| Jun-84 | SSC Central Design Group at LBL started | Maury Tigner |
| Nov-84 | Ronald Reagan reelected President | |
| Jan-87 | SSC site selection for the 87.1 km ring, 43 proposals | |
| Nov-88 | George H. Bush elected President, Texas SSC site chosen | |
| Aug-89 | LEP starts operation, 26.7 km ring | |
| Sep-89 | Berlin wall falls | |
| Dec-91 | Soviet Union disintegrates | |
| Jan-92 | UNK construction ends | Victor Yarba |
| Nov-92 | W. Clinton, A. Gore elected President and Vice-President | |
| Sept-93 | US/Russia agree to build Space Station | Gore/Chernomyrdin |
| Nov-93 | Texas SSC terminated | |
| Jun-99 | First Space Station module, Zarya launched | |
| Dec-00 | LEP shuts down to begin LHC construction | |
| Sept-08 | LHC starts, quench incident, LHC down | |
| Nov-09 | LHC resumes operation | |

**Table 1. Accelerator Time Lines**

## UNK

The UNK has a ring about three times larger than the Fermilab accelerator. Like the Tevatron UNK will contain a ring of conventional magnets to bring protons to 400 GeV. The superconducting ring will then accelerate protons to 3 TeV. Thus in many ways it will be a Tevatron scaled up by a factor of three. In 1985 I visited the construction site and the imposing tunnel shown in Figure 4. Figure 5a: Gennady Gurov stands before a conventional dipole; Figure 5b: the coil of a superconducting dipole.

I was impressed with the UNK project; the tunnel was steel lined. It was of generous proportions – I suspect that Bob Wilson would have objected to that – but servicing and upgrades would have been much easier. I recall commenting that they should keep to their tight schedule. They would only be able to keep the title of “highest energy’ for a short time since Fermilab would build the SSC with an aggressive construction schedule – like Bob Wilson!

The disintegration of the USSR in late 1991, with the ensuing economic crisis, forced the closing of the UNK project. Most of the tunnel and about 80% of the conventional accelerator had been installed, and a 25 superconducting magnet string had been successfully tested.

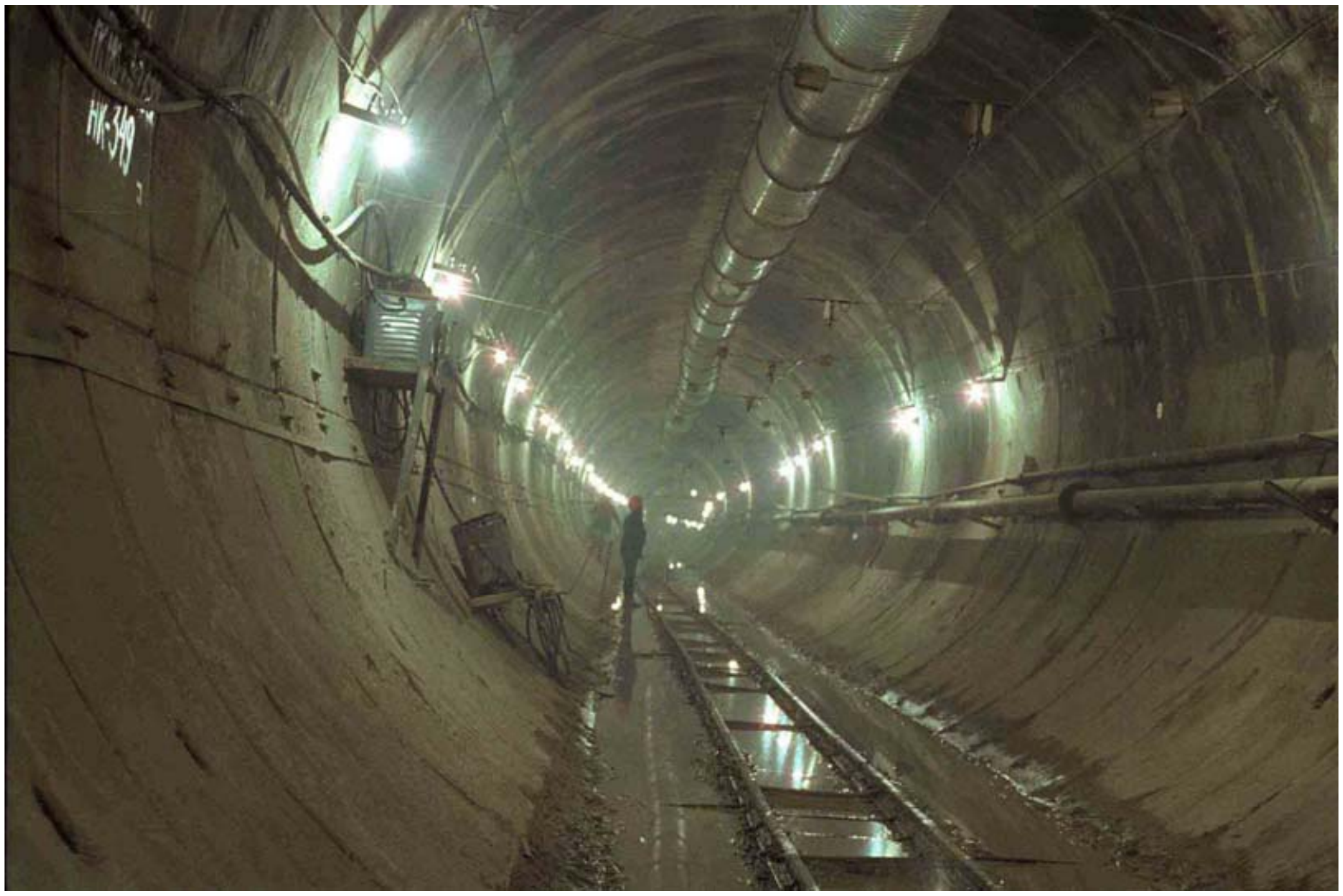

Figure 4, The UNK tunnel under construction

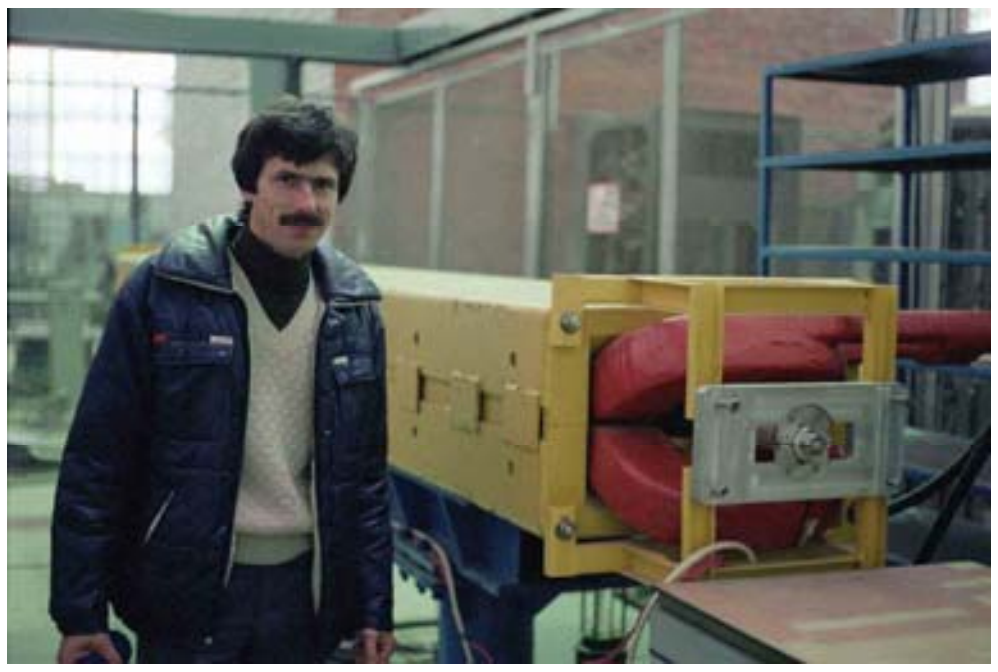

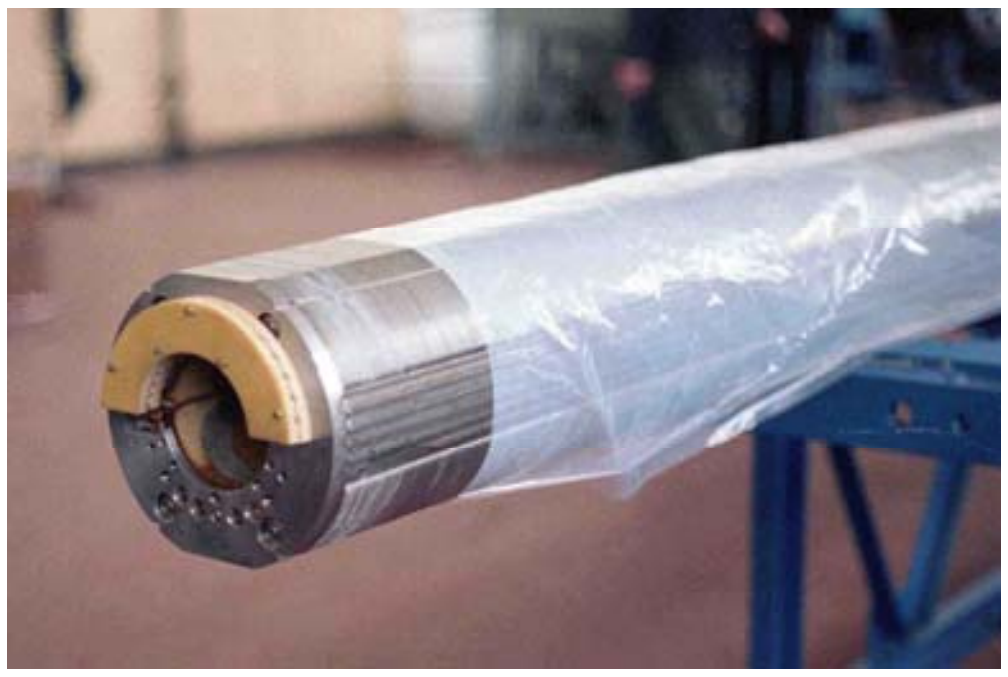

Figure 5a UNK conventional magnet. Figure 5b superconducting coil

## SSC

We had all heard about the SSC. We had been to workshops, summer studies, etc. helping to fix the parameters of such a machine. We were even more excited about the great physics that it would produce. Where would it be sited? Who would build it? Might it be at Fermilab? Those were the really big questions.

In the spring of 1983, Leon Lederman, then director of Fermilab, asked me to form a small group to address the question of whether Fermilab might be an appropriate site for the SSC? By then we knew it would be a 20 TeV accelerator with a circumference of about 52 miles. It certainly could not be contained within the Fermilab 6800 acre site. Could we go beyond the present site? What would the tunnel cost? Would it be on the surface like the Tevatron or a deeper tunnel? What was the geology like? Address the radiation safety problems, cost, location, etc.

The result was short paper that is the first entry in Table 1 and is reference 3. It is that document that led – about five years later - to the Illinois SSC proposal.

Within 50 miles of Fermilab is Chicago's Tunnel and Reservoir Plan (TARP). It is a water reclamation project that utilizes tunnels in dolomite rock under the Chicago region. Some of the almost 100 miles of tunnel are over 35 feet in diameter. This same rock stratum extends under Fermilab and the entire area of the proposed Illinois SSC site.

A major step, shown in Table 1, was the establishment of a Central Design Group at LBL headed by Maury Tigner. This led to a specific SSC design. I recommend[4] the excellent history of the Central Design Group by Stanley Wojcicki.

Continuing in Table 1 we see that in 1984 Ronald Reagan was elected to a second term and in his administration, the SSC site selection process was initiated. A total of 43 proposed sites were considered; a major national event patterned after the site selection process for Fermilab in 1967. A group, SSC For Illinois, Inc. was constituted to write the proposal. It would have been naïve to think that perception and politics would not be involved in the selection process. I recall a meeting where someone suggested that the major highway bordering Fermilab be named the "Ronald Reagan Memorial Highway" - because some other states that were making proposals were making similar changes. I-88 is still called the Ronald Reagan Memorial Highway!

In hindsight it should not have been a great surprise that Texas was chosen as the SSC site on the day that George H. Bush was elected president.

The new SSC site in Waxahachie Texas had many problems. There were organizational issues, major cost overruns, and stretched out schedules. These are described in a second article[5] by Stanley Wojcicki.

In 1992 W. Clinton was elected president. In 1993 the US and Russia signed an agreement to build the Space Station. This major and costly project of the new administration and the discouraging progress of the Texas SSC prompted Congress to terminate funding for the SSC in 1993.

Stanley Wojcicki[5]: " **In retrospect, choosing a green site in Texas was a mistake**".

This mistake cost billions of dollars and set high-energy physics back by at least a decade. **Why, who, and how was this mistake made**?

# LHC

In 1986 CERN was constructing LEP; the world's largest accelerator for the collisions of electrons and positrons. Its 26.7 km length would extend into France well beyond the borders of CERN's existing Swiss site. This site would be pushing the limits of such a ring since it would extend into the challenging geology of the Jura mountains. Figure 6 taken during the construction in 1986 shows water inflow that delayed construction but was eventually contained.

The successful LEP program (Table 1) began in 1989 and concluded at the end of 2000. An important decision was made to use the LEP tunnel for the LHC. That entailed a removal of the smaller LEP magnets and the installation of the much larger LHC magnets and the construction of the two large caverns to house the massive CMS and ATLAS detectors. This enormous project took about seven years to complete.

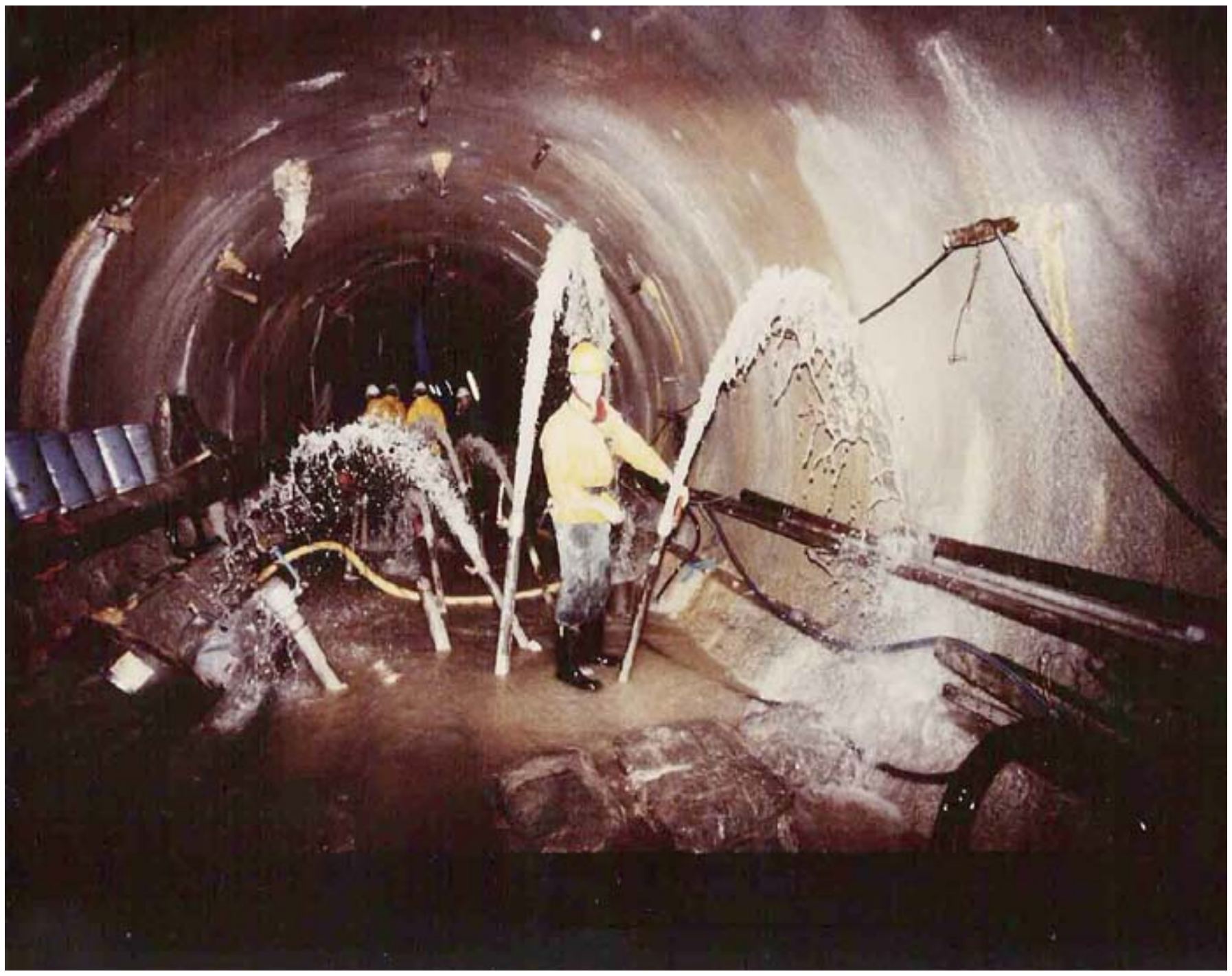

Figure 6. LEP tunnel water in the Jura region.

This new configuration for the LHC is shown in Figure 7. The proposed SSC ring size is a factor of 3.25 larger than the LHC ring so that even when the LHC is operating at full energy - 7.0 TeV - it would be about 1/3 of the SSC energy. To reach 7.0 TeV, the LHC magnets must operate at 8.3 T. The SSC magnets would operate at only 6.79 T at full energy. Thus the LHC requires a truly "heroic" cryogenic system. Figure 8, taken from the CERN web page "Facts and Figures" describes the magnitude of the cryogenic system.

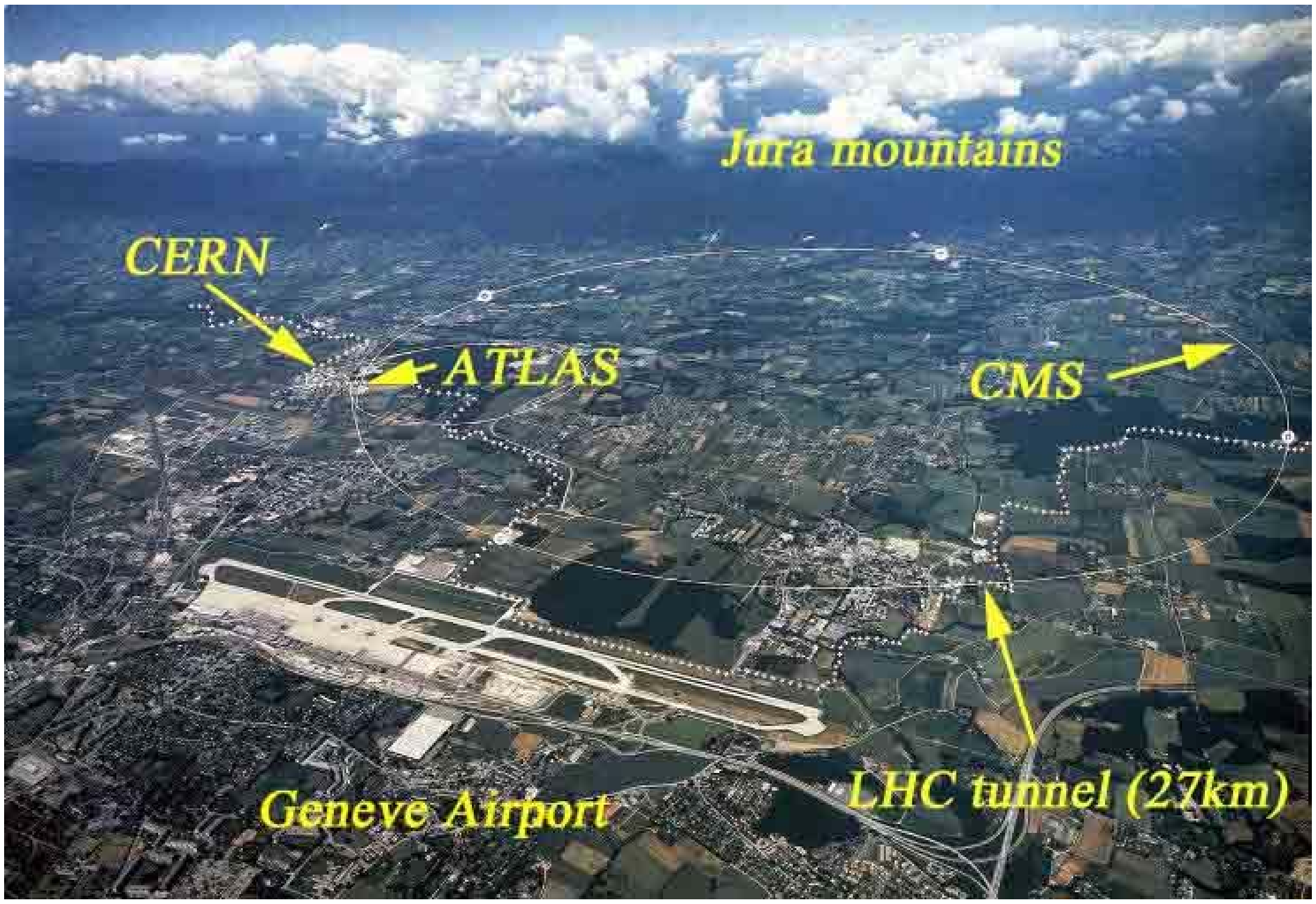

Figure 7. The LHC tunnel, detectors, and the Jura.

**The largest machine in the world...**

The precise circumference of the LHC accelerator is 26 659 m, with a total of 9300 magnets inside. Not only is the LHC the world's largest particle accelerator, just one-eighth of its cryogenic distribution system would qualify as the world's largest fridge. All the magnets will be pre-cooled to -193.2°C (80 K) using 10 080 tonnes of liquid nitrogen, before they are filled with nearly 60 tonnes of liquid helium to bring them down to -271.3°C (1.9 K).

Figure 8. **LHC cooling**

The LHC started operation in the September 2008 and suffered a major quench incident that damaged a couple dozen magnets and vented several tons of He. The LHC did not fail gracefully!

The recovery in late 2009 – at half energy – is proceeding cautiously. We hope and expect that the LHC will recover and become a robust, healthy and productive machine. With the loss of the SSC and UNK, the LHC remains our only hope for an accelerator beyond the Tevatron.

## ACKNOWLEDGMENTS

I am indebted to Victor Yarba for information about UNK; the photos are my own. Comments from Alexei Vorobyov on the history of UNK were appreciated.

Many people shared their thoughts on the Texas SSC including John Peoples, Chris Quigg, Victor Yarba, Chris Laughton, and Bruce Chrisman. I thank Stan Wojciciki for sharing with me his insightful history articles.

I thank Chris Laughton for use of his LEP tunnel picture. Other CERN pictures and words came from the CERN web site.

I thank Miguel A. Perez for inviting me to his meeting in Mexico City and pressing me to record my presentation. This is a fitting way to honor the memory of our friend and colleague Augusto Garcia.

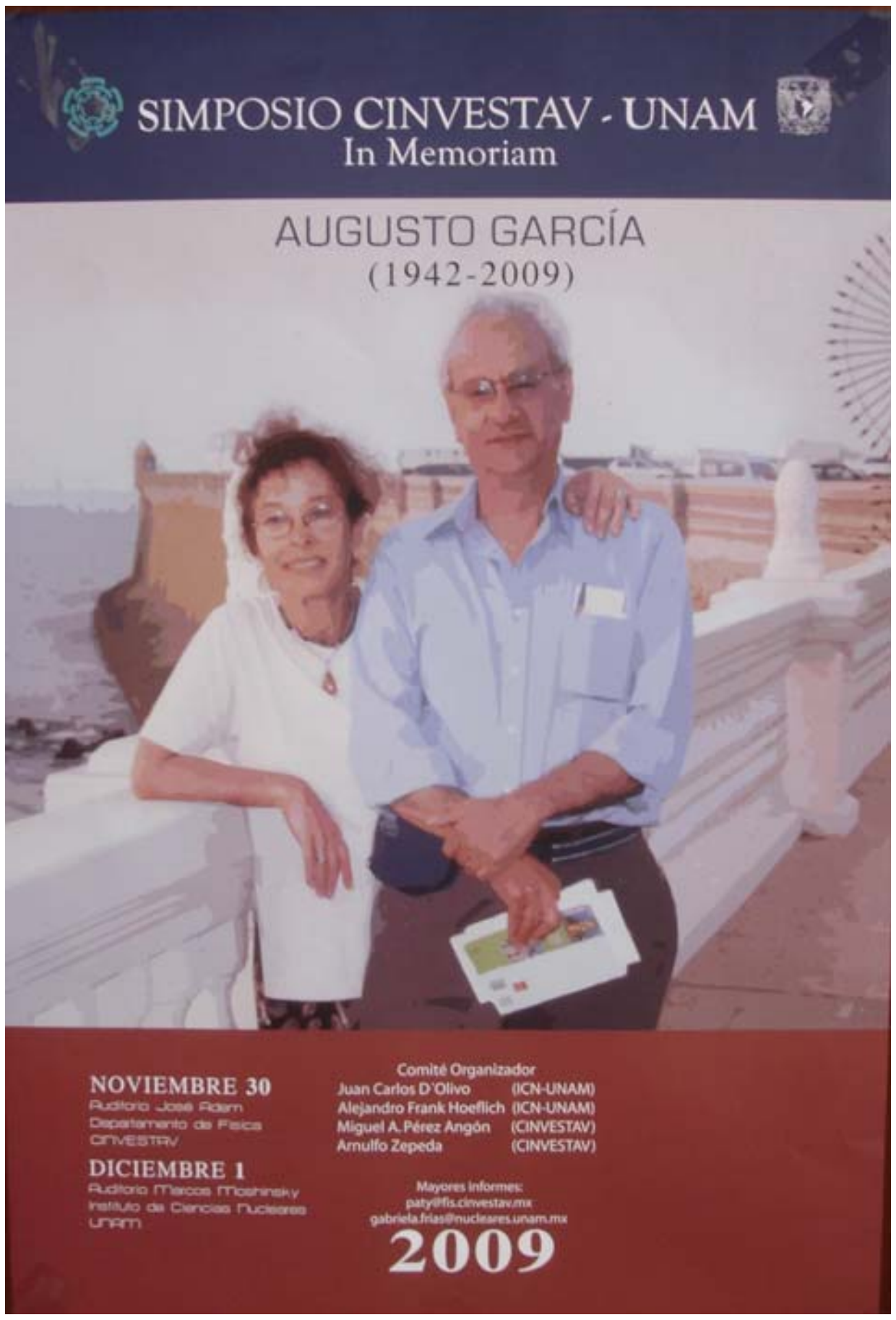

I also wish to thank Bob Wilson for teaching us how to build accelerators with beauty, imagination and economy.